%% file: main.tex
\documentclass[aip,reprint,onecolumn]{revtex4-1}
\usepackage{graphicx,amsmath,amssymb,float,geometry,chngcntr}
\input{defs}

\begin{document}
\linespread{1}
\counterwithout{equation}{section}

\title{Asymmetric Flexural-gravity Lumps in Nonuniform Media}

\author{Yong Liang}
\author{Mohammad-Reza Alam}
\email{reza.alam@berkeley.edu}
\affiliation{Department of Mechanical Engineering, University of California Berkeley, Berkeley, CA 94720}

\begin{abstract}
Here we show that asymmetric fully-localized flexural-gravity lumps can propagate on the surface of an inviscid and irrotational fluid covered by a variable-thickness elastic material, provided that the thickness varies only in one direction and has a local minimum. We derive and present equations governing the evolution of the envelope of flexural-gravity wave packets allowing the flexing material to have small variations in the transverse (to propagation) direction. We show that the governing equation belongs to the general family of Davey-Stewartson equations, but with an extra term in the surface evolution equation that accounts for the variable thickness of the elastic cover. We then use an iterative Newton-Raphson scheme, with a numerical continuation procedure via Lagrange interpolation, in a search to find fully-localized solutions of this system of equations. We show that if the elastic sheet thickness has (at least) a local minimum, flexural-gravity lumps can propagate near the minimum thickness, and in general have an asymmetric bell-shape in the transverse to the propagation direction. In applied physics, flexural-gravity waves describe for instance propagation of waves over the ice-covered bodies of water. Ice is seldom uniform, nor is the seafloor, and in fact near the boundaries (ice-edges, shorelines) they typically vary only in one direction (toward to edge), and are uniform in the transverse direction. This research suggests that fully localized waves are not restricted to constant ice-thickness/water-depth areas and can exist under much broader conditions. Presented results may have implications in experimental generation and observation of flexural-gravity (as well as capillary-gravity) lumps.

\end{abstract}
\maketitle

\vspace{-0.5cm}
\section{Introduction}

Fully localized solitary waves (and wavepackets) are of importance in applied sciences as they can transport mass, momentum and energy over long distances. It is known today that gravity waves cannot admit fully-localized waves\cite{Beals1986,Craig2002}, but in the presence of surface-tension or flexural-rigidity on the water surface such localized structures may exist\cite[e.g.][]{Fokas1990,Alam2013,Berger2000,Kim2005a,Parau2007b}. These structures are formed as a result of a careful balance between nonlinearity and the dispersion. 

Of interest in this paper, is the fully-localized wave packets of flexural-gravity systems. Equations governing the evolution of flexural-gravity wave packets are described by the Davey-Stewartson equation \cite{Alam2013,Milewski2013}. Davey-Stewartson (DS) equation \cite[][]{Davey1974} is a two-dimensional extension of Nonlinear Schr\"odinger (NLS) equation\cite{Hasimoto1972,Zakharov1968}. Compared to two-dimensional Nonlinear Schr\"odinger (2DNLS) equation, DS further includes the coupling with an auxiliary field (which is the mean current in the context of hydrodynamics). In contrast to one-dimensional NLS, 2DNLS is not integrable and does not admit localized solutions. The coupling with the mean field allows finite-depth DS equation to be integrable (under certain conditions), and to admit fully-localized solutions. In the limit of infinite depth DS equation reduces to 2DNLS.  

While DS equation has been studied extensively due to its importance in different areas of science  \cite[e.g.][]{Champagne1988,Clarkson1994,Li2008,Hizel2009,GUndersen1993}, investigation of its fully-localized solutions is a relatively young field of research. Two families of such solutions have so far been discovered: Dromions and Lumps. Dromions are fully-localized surface structures with exponentially decaying tails that form at the intersection of mean-flow line-solitary tracks \cite{Boiti1988,Djordjevic1977,Fokas1990,hirota2004,Gilson1991,Hietarinta1990}. Lumps on the other hand have algebraically decaying tails but both the surface elevation and the mean field are fully-localized \cite{Satsuma1979a,Kim2005}. Lump solution is not restricted to DS, but several other systems including Kadomtsev-Petviashvili equation \cite[][]{Manakov1977,Ablowitz1979}, Benney-Luke equation \cite[][]{Berger2000} and full Euler equation \cite[][]{Groves2008,Parau2005} also admit lump solutions. 

Flexural-gravity waves in two-dimension have been the subject of considerable interest from various standpoints \cite[e.g.][]{Vanden-Broeck2011,Toland2007,Forbes1986,Forbes1986b,Wang2013}. In three-dimensional few existing research endeavors are mainly focused on the problem of  moving loads on the ice \cite[e.g.][]{Parau2002,Miles2003,Bonnefoy2009,Parau2011,Davys1985,Schulkes1987,Guyenne2014}. The first consideration of fully-localized waves in three-dimension, to our knowledge, has been made recently where it has been shown that in a constant-depth water, wave packets of flexural-gravity waves may admit three-dimensional fully-localized structures in the form of lumps \cite[][]{Milewski2013,Wang2014} and dromions \cite[][]{Alam2013,Liang2013}.

Here we consider weakly nonlinear flexural-gravity wave packets propagating over a (relatively) thin but variable-thickness elastic cover that lies on the surface of an inviscid and incompressible fluid. A perturbation scheme is used to derive the evolution equation for the envelope of waves, and it is shown that the governing equation belongs to the general family of Davey-Stewartson \cite[][]{Davey1974,Milewski2013,Alam2013}, but with an extra term (linear in the surface elevation) whose coefficient is a function of thickness variations.

We consider the Elliptic-Elliptic subset of this equation for which, in the absence of thickness variation, lump solution exists. We then look for fully-localized lump solutions for a variety of thickness variation functions with the help of Lagrange interpolation of the lump shape combined with Newton-Raphson iteration scheme to solve the equation through a numerical continuation procedure without prescribed boundary conditions. We show that lump solution over a transversely variable thicknesses exists if the thickness function has a local minimum. These lumps, however, have an asymmetric shape in the transverse to the propagation direction with a steeper slope on the side that the gradient of thickness variation is higher. For a constant thickness and variable depth, the governing equation is similar in the form, but different in the coefficient. In this case lump solutions exist if the depth perturbation has a maximum, or in other words if the water depth has a minimum. In the nature and in practice, media through which waves propagate is usually non-uniform. Results presented here show that fully-localized solutions are not limited to perfectly uniform environments and in fact can exist and propagate in media with variable properties. 

\vspace{-0.5cm}
\section{Governing Equation}
Consider an incompressible, inviscid and homogeneous fluid of density $\rho_f$ and constant depth $h$ bounded on top by a thin sheet of an elastic material (such as ice) with the density $\rho_{i}$ and spatially variable (small) thickness $L(x,y)$. We define a Cartesian coordinate system with the $x,y$-axes on the mean bottom and $z$-axis positive upward. Assuming that flow is irrotational, a potential function $\phi$ can be defined such that $\vec \nabla\phi\equiv\vec u$ where $\vec u=\vec u(x,y,z,t)$ is the Eulerian velocity of the flow field. If $\eta=\eta(x,y,t)$ denotes the elevation of the free surface from the mean water level, the governing equations read
\bes\label{910}
&&\phi_{xx}+\phi_{yy}+\phi_{zz}=0,~~0<z<h+\eta(x,y)\\
&&\phi_{z}=\eta_{t}+\phi_{x}\eta_{x}+\phi_{y}\eta_{y},~~z=h+\eta(x,y)\\
&&\phi_{t}+\frac{1}{2}(\phi_{x}^{2}+\phi_{y}^{2}+\phi_{z}^{2})+g\eta=
-\lb H\nabla^4 \eta + 2 H_{x}\nabla^2\eta_{x}+2H_{y}\nabla^2\eta_{y}+\nabla^2 H\nabla^2 \eta\right. \nn \\
&&\left.-(1-\nu)(H_{xx}\eta_{yy}-2H_{xy}\eta_{xy}+H_{yy}\eta_{xx})+R\eta_{tt}\rb, ~~~~z=h+\eta(x,y)\label{911}\\
&&\phi_{z}=0,~z=0 \label{989}
\ees
where $H=EL^3/12\rho(1-\nu^2),~R=\rho_{i}L/\rho_{f}$ \cite[c.f. e.g.][]{Byers2013}. We define the following dimensionless variables
\be\label{740}
&&x^*,y^*=\f{x,y}{\lambda},~~~z^*=\f{z}{h},~~~\eta^*=\f{\eta}{a},~~~t^*=\f{t\sqrt{gh}}{\lambda}\nn\\
&&\phi^*=\f{\phi h}{\lambda a \sqrt{gh}},~H^*=\f{H}{g\lambda^4},~~~R^*=\f{R h}{\lambda^2},~\ep=\f{a}{h},~\delta=\f{h}{\lambda},
\ee
where $\lambda$ is the characteristic wavelength of the carrier wave and $a$ is the characteristic amplitude. Using definitions Eq. \eqref{740}, the dimensionless form of the governing equation Eq. \eqref{910}, after dropping asterisks for notational simplicity, becomes
\bes
&&\phi_{zz}+\delta^2(\phi_{xx}+\phi_{yy})=0,~~0\leq z\leq 1+\epsilon \eta,\\
&&\phi_{z}=\delta^2(\eta_{t}+\epsilon \phi_{x}\eta_{x}+\epsilon \phi_{y}\eta_{y}),~~z=1+\epsilon \eta,\\
&&\phi_{t}+\frac{1}{2}\epsilon (\phi_{x}^2+\phi_{y}^2+\frac{1}{\delta^2} \phi_{z}^2)+\eta=
-\lb H\nabla^4 \eta + 2 H_{x}\nabla^2\eta_{x}+2H_{y}\nabla^2\eta_{y}+\nabla^2 H\nabla^2 \eta\right. \nn \\
&&\left.-(1-\nu)(H_{xx}\eta_{yy}-2H_{xy}\eta_{xy}+H_{yy}\eta_{xx})+R\eta_{tt}\rb, ~~z=1+\epsilon\eta\slabel{9112},\\
&&\phi_{z}=0,~~z=0.\slabel{988}
\ees

We are interested in weakly nonlinear harmonic waves of wavenumber $k$ with slowly varying amplitudes in both $x,y$ directions. To achieve this solution we assume $\ep\ll O(1)$ but leaving $\delta$ to be arbitrary. 

Permanent-form fully-localized structures, for example lumps, are not expected to exist if the medium (e.g. thickness of the elastic cover) changes in the direction of the propagation (in this paper we take the direction of the motion along the $x$-axis). In fact, if the medium has a constant mean but with random perturbations, we expect that the waves attenuate over time, as they propagate, via the so-called localization mechanism \cite[][]{Belzons1988,Devillard1988}. Therefore here we assume that the thickness of the elastic cover only changes in the transverse (i.e. $y$) direction and also very slowly, that is, $L(x,y)=L(y)=L_0[1+\epsilon^2 f(y)]$, where $L_0$ is the mean thickness and $f(y)$ is a continuous function that specifies how thickness varies about the mean and satisfies $f(y)\sim O(1),~f'(y)\sim O(\epsilon)$. Under these assumptions, the dynamic boundary condition Eq. \eqref{9112}, correct to the order $\epsilon^2$, becomes
\be
\phi_{t}+\frac{1}{2}\epsilon(\phi_{x}^{2}+\phi_{y}^{2}+\frac{1}{\delta^2}\phi_{z}^{2})+\eta=-H_0[1+3\epsilon^2 f(y)]\nabla^{4}\eta-R_0[1+\epsilon^2 f(y)]\eta_{tt}.
\ee
where $H_0=EL_0^3/12\rho (1-\nu^2)g\lambda^4$ and $R_0=\rho_{i}L_0h/\rho\lambda^2$. We further define the following different scale variables
\be\label{913}
\xi=x-c_pt,~~~\zeta=\ep(x-c_gt),~~~Y=\ep y,~~~\tau=\ep^2 t,
\ee
where $c_p(k),c_g(k)$ are respectively the phase and group velocity of the carrier wave. Note that $c_p(k),c_g(k)$ are unknown speeds at this stage, but leading order (i.e. linear) analysis proves $c_p(k)$ to be the phase velocity and first order analysis shows that $c_g(k)$ must in fact be the group velocity of the wave system. 

We assume that the solution to the governing equations can be expressed by a convergent asymptotic series in terms of our small parameter $\ep$. In terms of new variables Eq. \eqref{913} we suggest the series solution in the form \cite[c.f.][]{Davey1974}

\be
&&\phi(\xi,\zeta,Y,z,\tau)=f_0(\zeta,Y,\tau)+\sum_{n=0}^{\infty}\ep^n\lcb \sum_{m=0}^{n+1} F_{nm}(z,\zeta,Y,\tau)E^m+{\rm c.c.} \rcb,\hspace{1cm}\\
&&\eta(\xi,\zeta,Y,\tau)=\sum_{n=0}^{\infty}\ep^n\lcb \sum_{m=0}^{n+1} A_{nm}(\zeta,Y,\tau)E^m+{\rm c.c.} \rcb,\hspace{0.6cm}
\ee
where $E=\exp(ik\xi)$ and $A_{00}=0$. Leading order (i.e. linear) analysis provides the expression for the dispersion relation 
\begin{equation}
c_{p}^2=\frac{(1+\tilde{H})\tanh\delta k}{\delta k+\tilde{R}\tanh\delta k}.
\end{equation}
where $\tilde{H}=H_0k^4$ and $\tilde{R}=R_0k^2$.

If the perturbation analysis perused to the second order of nonlinearity, equations governing the evolution of the envelope  $A_{01}(\zeta,Y,\tau)$ and the mean field $f_{0}(\zeta,Y,\tau)$ obtain in the following form \cite[the procedure of arriving at the following equations are algebraically involved, but standard and can be found  in e.g.][]{M.RezaAlam2012,Johnson1997}

\bes\label{930}
&&\alpha f_{0\zeta\zeta}+f_{0YY}=-\beta |A_{0}|^2_{\zeta}\\ \slabel{930a}
&&iA_{0\tau}+\vartheta A_{0\zeta\zeta}+\mu A_{0YY}=(\nu_{1}|A_{0}|^2+\nu_{2}f_{0\zeta}+\nu_{3})A_{0}\slabel{930b}
\ees
in which
\bes
&&\alpha=1-c_{g}^2,~~\beta=\frac{1}{\sigma^2}[2\delta k c_{p}\sigma+(\delta^2k^2c_{p}^2c_{g})(1-\sigma^2)]\geq 0\nn\\
&&\vartheta =\frac{\omega''}{2},~~\mu=\frac{c_{g}}{2k}=\frac{\omega'}{2k}\geq 0\nn\\
&&\nu_{1}=\frac{k^3\delta}{4\sigma \omega}\Gamma,~~\nu_{2}=k[1+\frac{\delta^2k^2c_{p}c_{g}(1-\sigma^2)-2\tilde{R}\sigma^2}{2\sigma(\delta k +\tilde{R}\sigma)}]\geq 0\nn\\
&&\nu_{3}= f(y)\frac{2\omega(3\tilde{H}-c_{p}^{2}\tilde{R})\sinh\delta k}{\sinh\delta k(1+\tilde{H}-c_{p}^{2}\tilde{R})+c_{p}^{2}\delta k\cosh\delta k+2\tilde{R}c_{p}^{2}\sinh\delta k}\nn
\ees

where $\Gamma=p/q$ and
\bes
&&q= (\tilde{R}\sigma+\delta k)^3[(\tilde{R}\sigma+\delta k)(-3+12\tilde{H})+\delta k(1+\tilde{H})(3-\sigma^2)]\nn\\
&&p = a+b\sigma+c(1-\sigma^2)+d(1-\sigma^2)\sigma+e(1-\sigma^2)^2+f(1-\sigma^2)^2\sigma+g(1-\sigma^2)^3\nn\\
&&a = (52\tilde{H}^2+44\tilde{H}-8)\delta^4k^4+(48\tilde{H}^2+36\tilde{H}-12)\tilde{R}^2\delta^2k^2\nn\\
&&c = (-104\tilde{H}+8-112\tilde{H}^2)\delta^4k^4+(36-144\tilde{H}^2-108\tilde{H})\tilde{R}^2\delta^2k^2\nn\\
&&b = (100\tilde{H}^2+80\tilde{H}-20)\tilde{R}\delta^3k^3,~~d= (32-176\tilde{H}-208\tilde{H}^2)\tilde{R}\delta^3k^3\nn\\
&&e = (-42\tilde{H}-63\tilde{H}^2-28\tilde{H}^3-7)\delta^4k^4+(72\tilde{H}-24+96\tilde{H}^2)\tilde{R}^2\delta^2k^2\nn\\
&&f = (-30\tilde{H}-24\tilde{H}^3-51\tilde{H}^2-3)\tilde{R}\delta^3k^3,~~g = (-2\tilde{H}^3-6\tilde{H}^2-6\tilde{H}-2)\delta^4k^4\nn
\ees
in which $\sigma=\tanh\delta k$. Note that $f(y)$ is now in the coefficient $\nu_3$.

Eq. \eqref{930} is a more general form of Davey-Stewartson\cite{Davey1974} (also known as Benney-Roskes-Davey-Stewartson equation \cite{Benney1969,Milewski2013}). In the special case of Eq. \eqref{930} when the thickness of the top elastic layer does not vary, i.e. $f(y)=0$, and if its inertia can be neglected, i.e. $R_0=0$, then we recover Eq. (2.10) in Ref. \onlinecite{M.RezaAlam2012} or Eqs. (15),(16) in Ref. \onlinecite{Milewski2013}. The effect of inertia appears in all coefficients in Eq. \eqref{930}.

We finally note that the effect of variable thickness appears only in the new term $\nu_3A_0$ in Eq. \eqref{930b}. In order for underlying assumptions of Eq. \eqref{930} to be valid the function $f(y)$ must satisfy $|f(y)|=O(1)$, $|f'(y)|=O(\epsilon)$. 

\section{Numerical Method}

\begin{figure}
\centering
  \includegraphics[scale=0.52]{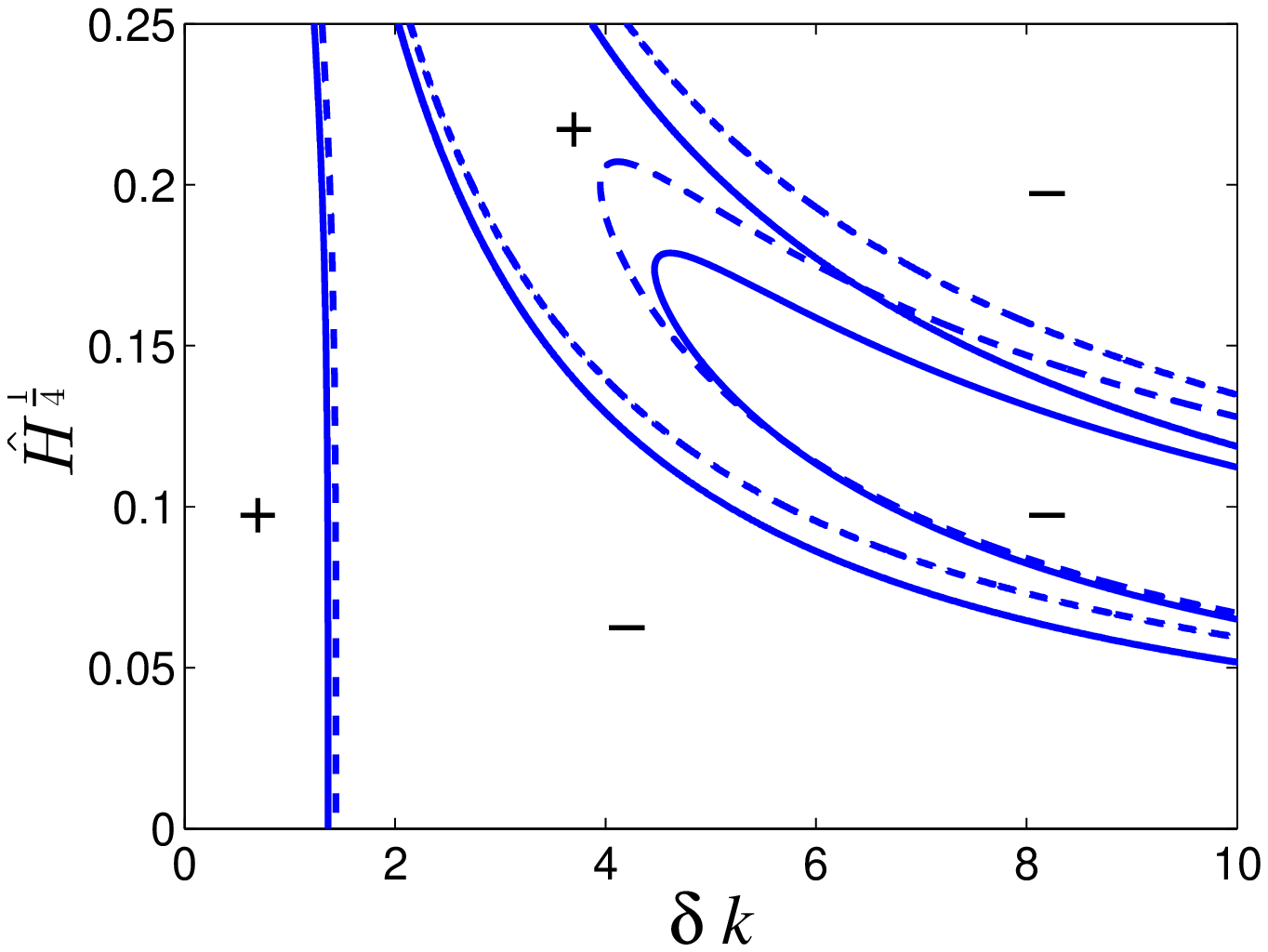}
	\put(-210,160){(a)}  
  \includegraphics[scale=0.52]{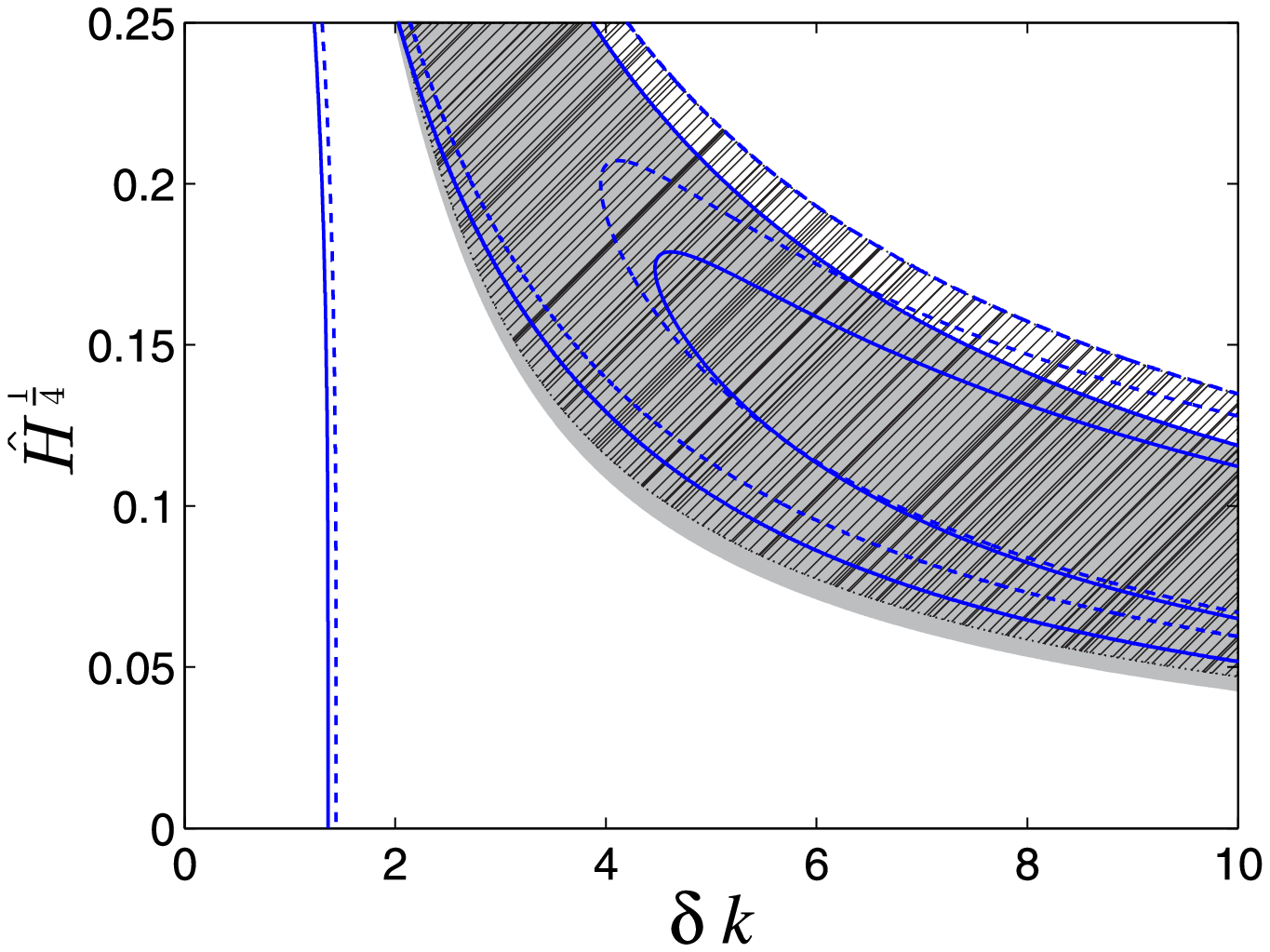}
	\put(-200,160){(b)} 
	\caption{(a) Focusing (denoted by ``+") and defocusing (denoted by ``-") areas of Davey-Stewartson equation with the effect of inertia of the elastic sheet ($\hat{R}=0.05$, dashed-line) and without the effect of inertia ($\hat{R}=0$, solid lines). The choice of $\hat{R}=0.05$ is made so that the figure shows clearly direction of departure from the ground state of $\hat{R}=0$. (b) Areas of elliptic-elliptic Davey-Stewartson for $\hat{R}=0.05$ (hatched area) and $\hat{R}$=0 (gray area).}\label{Rhat}
\end{figure}

To cast Eq. \eqref{930} in a more commonly-used form we define $v=-f_{0\zeta}+g|A_0|^2$, where $g=-\beta/\alpha$ and for simplicity of notations replace $A_0,\zeta,Y$ with $u,x,y$. Eq. \eqref{930} turns into
\bes\label{625}
&&\alpha v_{xx}+v_{yy}=g|u|^2_{yy}\\
&&iu_{\tau}+\vartheta  u_{xx}+\mu u_{yy}+\nu_{2}uv+\gamma|u|^2u-\nu_{3}u=0
\ees
where $\gamma=-\nu_{1}-\nu_{2}g$. Depending on whether $\gamma>0$ or $\gamma<0$ Davey-Stewartson equation is known as, respectively, focusing or defocusing. Likewise signs of $\alpha$ and $\vartheta$ determine whether Eq. \eqref{930} is elliptic-elliptic, elliptic-hyperbolic, hyperbolic-elliptic or hyperbolic-hyperbolic (the latter is not possible in the context of water waves\cite{Cipolatti1992}). Figure \ref{Rhat}(a) compares the area of focusing and defocusing of the governing equation Eq. \eqref{930} for $\hat{R}\equiv\tilde{R}/(k\delta)^2=$0, 0.05. The new variable $\hat R$ in physical space is simply the ratio of the mass of the elastic sheet to the mass of the water underneath it, and therefore is independent of the wavelength. Whether Eq. \eqref{930} is focusing or defocusing is a function of both $k\delta$ and $\hat{H}\equiv\tilde{H}/(k\delta)^4$ and this functionality is shown in Figure \ref{Rhat}(a). Areas for which the governing Davey-Stewartson is elliptic-elliptic is overlain on top of boundaries of focusing/defocusing curves in Figure \ref{Rhat}(b). The case of $\hat{R}$=0 is shown by gray area, and for $\hat{R}$=0.05 by hatched area.

Here we only consider focusing elliptic-elliptic Davey-Stewartson equation which is known to admit localized solitary waves\cite{Cipolatti1992}. We further assume that the amplitude has a periodic phase with an unknown frequency $\Omega$, i.e. 

\be\label{334}
u=r(x,y) e^{i\Omega\tau}.
\ee

By implementing the following re-scaling 
\be
\bar{r}=r \Omega^{-1/2}\gamma^{1/2},~\bar{v}=\f{\gamma}{g\Omega}v,~\bar{x}=x\sqrt{\frac{\Omega}{\vartheta }},~\bar{y}=y\sqrt{\frac{\Omega}{\mu}}\nn
\ee
After dropping bars Eq. \eqref{625} becomes
\bes\label{940}
&&\mu\alpha\vartheta ^{-1}v_{xx}+v_{yy}=|r|^2_{yy}\slabel{941}\\
&&-r+r_{xx}+r_{yy}+\nu_{2}g\gamma^{-1}r v+r^3-\nu_{3}\Omega^{-1}r=0\slabel{942}.
\ees

The objective here is to look for lumps solutions of Eq. \eqref{940} particularly when $f(y)$ is nonzero and non-constant (note that $f(y)$ is hidden in $\nu_3$). The numerical procedure is as follows: we first consider Eq. \eqref{942} with $\nu_2=\nu_3=0$, i.e., $-r+r_{xx}+r_{yy}+r^3=0$. This equation can be solved numerically using shooting method under the condition that $r(x,y)$ is axisymmetric and vanishes at infinity. It has countably many solutions, and the one that is always positive and asymptotically goes to zero as $r\rightarrow \infty$ (the so-called ground state) will be considered here as a first guess for solving Eq. \eqref{940}\cite[c.f.][]{Kim2006a,Milewski2013}. 

With the ground state solution in hand, we aim at solving   Eqs. \eqref{941} and \eqref{942} together including $\nu_2$ but keeping $\nu_3=0$. It is to be noted that Eq. \eqref{940} with $\nu_3=0$ is invariant under the parity transformation (i.e. when $x \rightarrow -x$, likewise for $y$). Although this by itself does not guarantee that the solution is parity invariant, we restrict our attention to solutions of Eq. \eqref{940} (with $\nu_3=0$) that are symmetric about both $x$- and $y$-axis \cite[c.f.][]{Milewski2013,Kim2005}. This assumption  later on will be relaxed in the $y$-direction in a search for asymmetric lumps. We express $r(x,y)$ and $v(x,y)$ in terms of Lagrangian polynomials whose coefficients are found via the iterative Newton-Raphson method. Since lumps have algebraically decaying tails the transformation $x,y=Lp,q/\sqrt{1-(p,q)^2}$ is used to map physical domain ($-\infty, \infty$) to a finite computational domain(-1,1). The effect of the right-hand side of Eq. \eqref{941} is included via numerical continuation (i.e. by slowly increasing an artificial coefficient in front of that term from zero to one \cite[][]{Kim2006a}). 

To find lump solutions of Eq. \eqref{940} including $\nu_3$ term, we start from the symmetric lump of the previous step and implement another level of numerical continuation on the $\nu_3$ term. Note that at this stage symmetricity in $y$-direction is relaxed. 

\section{Results and Discussion}

We show here that a variable-thickness elastic sheet, with the thickness variation function $f(y)$,  overlying the surface of an inviscid and irrational fluid can admit asymmetric flexural-gravity lumps. 

We first consider a case with the thickness variation function $f(y)=5\text{exp}(-y^2)\sin y$. Chosen parameters are chosen as  $\delta k$=2.8, $\hat{R}$=0.01,  $\hat{H}^{\f{1}{4}}$=0.19, $\Omega$=1/3, and domain transformation variable $L$=5. The number of grid points in the simulation domain is set at 24$\times$96. Note that computational domain is half of the full domain. Therefore this choice of grid size implies that the full domain is 48$\times$96. Chebyshev nodes are used as grid points to reduce the effect of Runge phenomenon. Targeted computational error is set to $1e-12$. It is to be noted that for the same parameters different choices of $L$ result in different lump solutions. This is not unexpected as, for instance, analytical dromion solution to Davey-Stewartson equation has several free parameters and for every set of parameters an infinite number of dromions can be found \cite[see e.g.][]{Gilson1991,M.RezaAlam2012}. There are, however, limitations on the choice of this number because for smaller $L$ we get a finer mesh close to the center of the domain of computation and coarser grid farther away and vice versa. Therefore care must be taken in a proper choice of scaling parameter $L$. 

\begin{figure}
  {\includegraphics[scale=0.5]{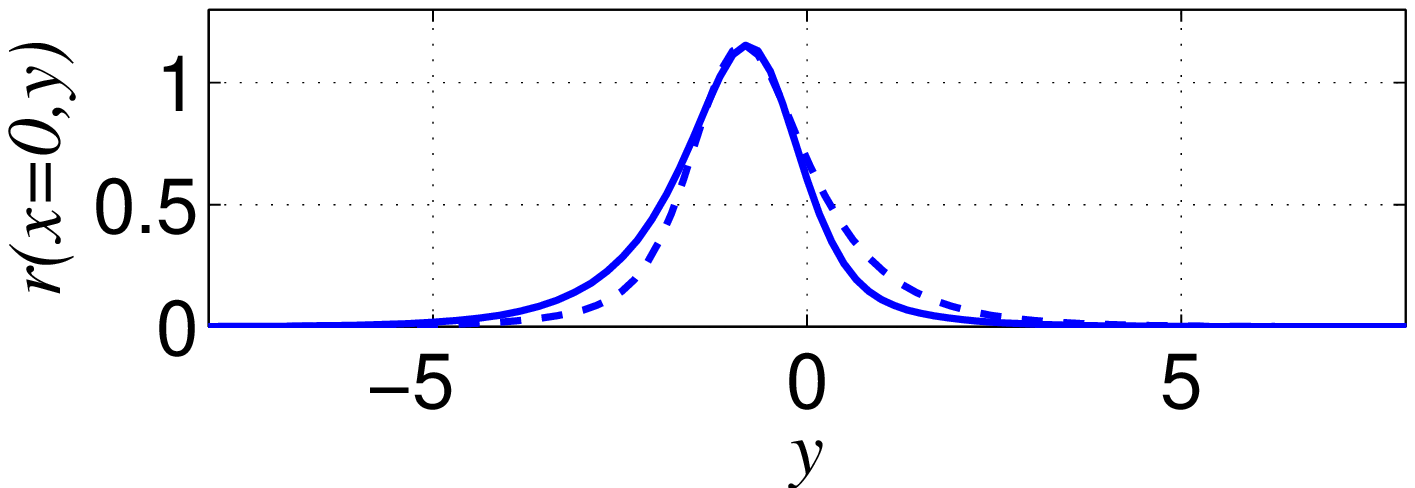}}\quad
  \put(-165,55){(a)}    \quad
  {\includegraphics[scale=0.5]{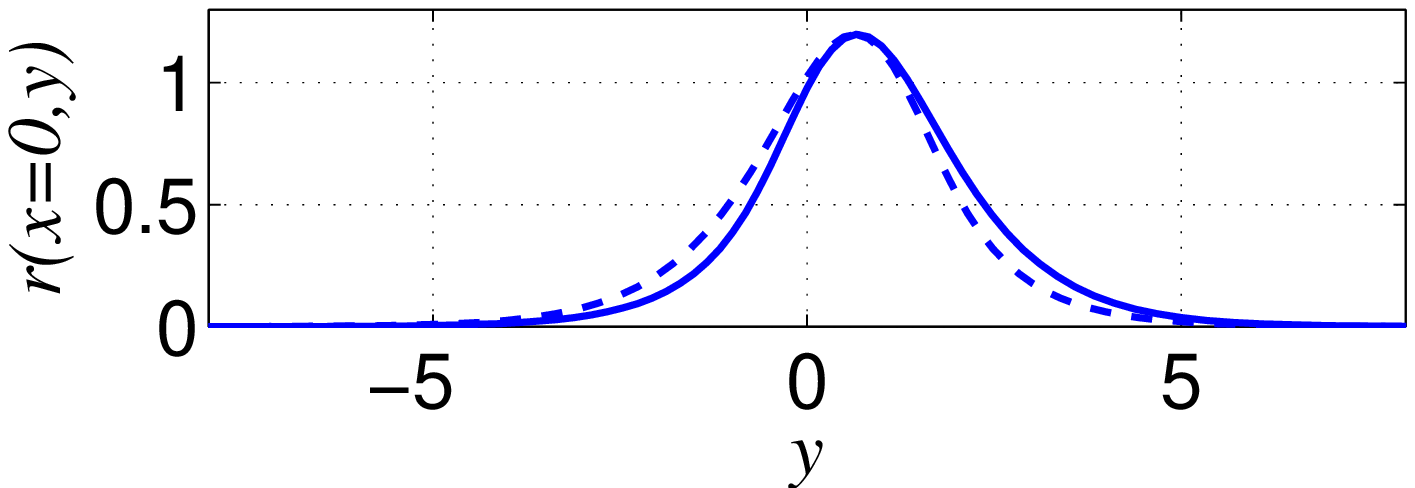}}\quad
  \put(-165,55){(d)}    \quad
	{\includegraphics[scale=0.5]{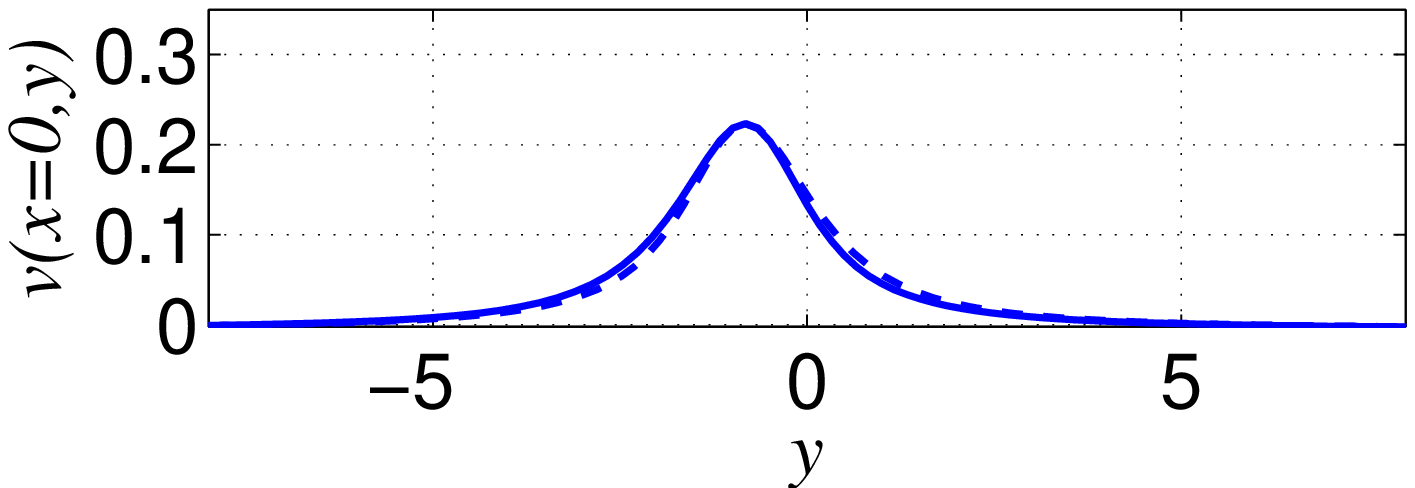}}\quad
  \put(-165,55){(b)}    \quad
	{\includegraphics[scale=0.5]{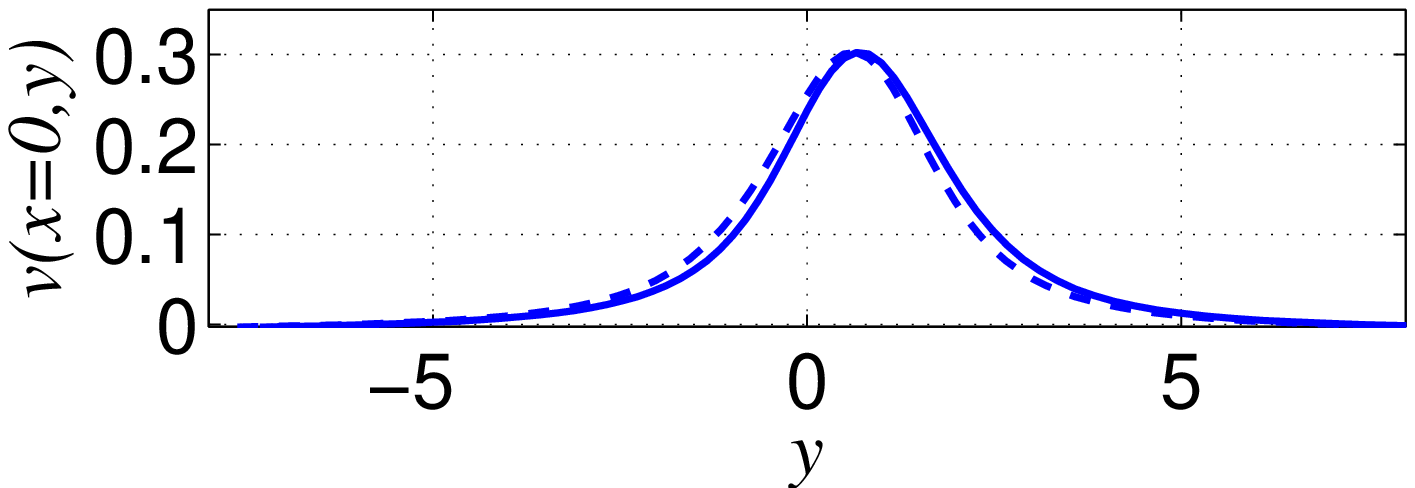}}\quad
  \put(-165,55){(e)}    \quad
  {\includegraphics[scale=0.5]{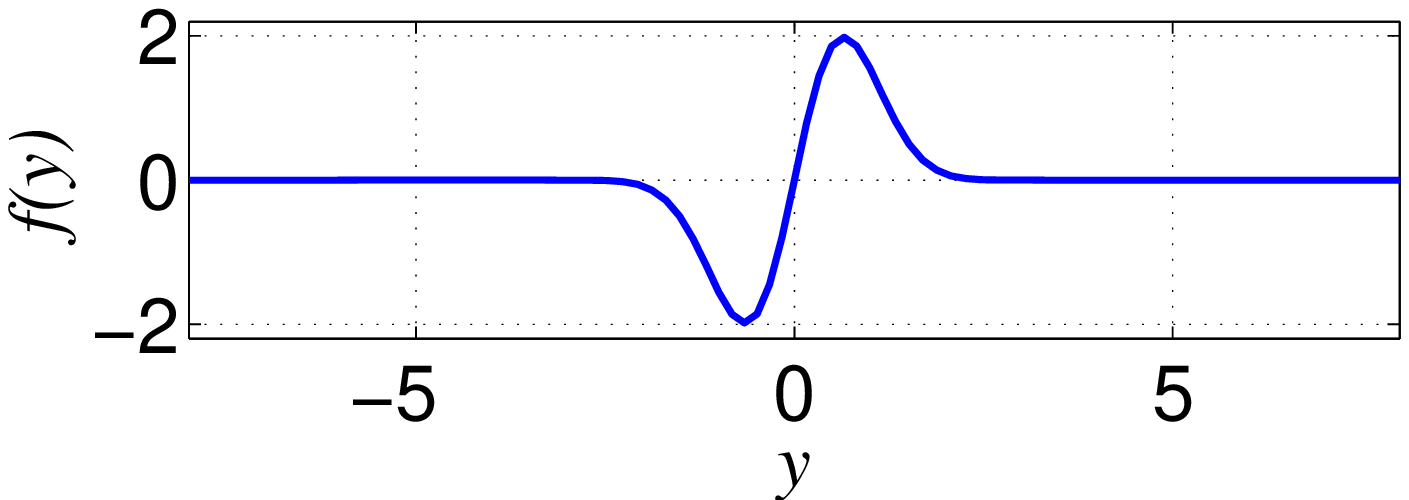}}\quad
  \put(-165,55){(c)}    \quad
	{\includegraphics[scale=0.5]{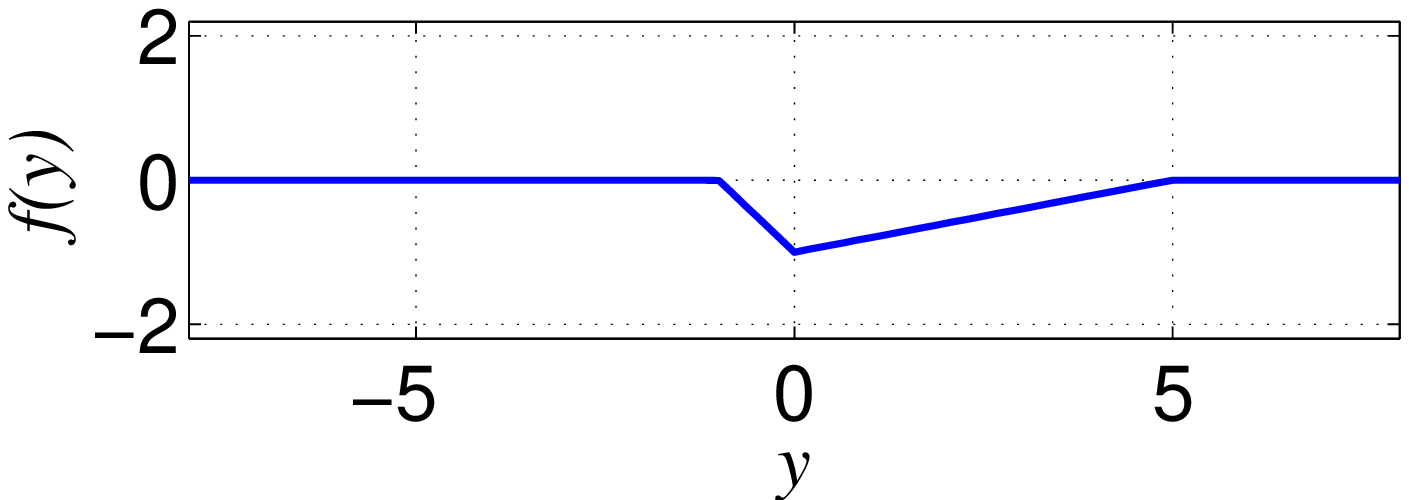}}\quad
  \put(-165,55){(f)}    \quad
  \caption{Cross sections of asymmetric flexural-gravity lumps propagating over a transversely uneven elastic sheet. (a),(b) cross sections $r(x=0,y)$, $v(x=0,y)$ for an asymmetric lump propagating over an elastic sheet with the thickness variation function  $f(y)=5\exp (-y^2) \sin y$ in (c). Dashed lines show the mirror of the right/left side profile (about the peak) on the left/right side and are presented to highlight the asymmetry of the shape. (d),(e) cross sections for the thickness variation function $f(y)$ in (f)(as is shown in Eq. \eqref{1243}). Parameters used here are $\delta k=2.8,~\hat{R}=0.01, \hat{H}^{\f{1}{4}}=0.19$, $\Omega=1/3$ and $L=5$.}\label{case1}
\end{figure}

Figures \ref{case1}(a)-(c) show the thickness variation function $f(y)$ (Figure \ref{case1}(c)) along with central cross-sections of $r,v$, i.e. $r(x=0,y)$ (Figure \ref{case1}(a)) and $v(x=0,y)$ (Figure \ref{case1}(b)). To highlight that the lump is asymmetric, the mirror image of each side about the peak of the lump is plotted by dashed-lines on the other side of the peak. In the absence of the thickness variation (i.e. when $f(y)$=0), the peak is at y=0. With the current form of $f(y)$ the peak is slipped away from $y$=0 toward the trough of the thickness variation function. A close examination of the Figures \ref{case1}(a)-(c) shows that the peak of the lump is not exactly at the trough but slightly to the left (this is further highlighted in the next example). The lump profile is leaned forward toward the steeper side of $f(y)$, i.e., it is steeper on the side where $f(y)$ is steeper.

To further highlight the effect of steepness of the profile on the shape of the lumps we consider a piecewise linear thickness variation function given by (see Figure \ref{case1}(f))
\be\label{1243}
f(y)=\begin{cases}
-y-1 & -1 <y< 0 \\
\f{1}{5}y-1& 0 <y< 5 \\
0 & \text{otherwise.}
\end{cases}
\ee

Note that Eq. \eqref{1243} has only one trough (minimum) and no distinguished crest (maximum). With the choice of parameters the same as in the previous case, the lump surface and associated mean current are shown in Figures \ref{case1}(d)-(e). Clearly the lump and the mean current have a steeper slope on the side that $f(y)$ is steeper. And the crest of the lump is formed away from the trough of the thickness on its right where the slope is milder. The trends are similar to the previous case but further highlighted. Note that strictly speaking a thickness variation function must be at least two times differentiable (i.e. $f(y)\in \mathcal C^2$, c.f. Eq. \eqref{911}), but nevertheless no derivative of $f(y)$ appears in the final form of governing equation Eq. \eqref{940}. Our numerical experiments show that the obtained asymmetric shape for the lump is quite robust to the perturbations to the thickness variation function. The last case of our interest is if $f(y)$ has a crest, but no trough, e.g. mirror of $f(y)$ shown in Figure \ref{case1}(f) about the horizontal axis. In this case our iterative algorithm does not converge to a lump solution. Iteration of the shape of the initial profile shows that the peak of the lump slips away from the crest of the topography (perhaps in a search for a trough), but continues to move to the right or left and never reaches a convergence. 

\begin{figure}
\centering
  {\includegraphics[scale=0.6]{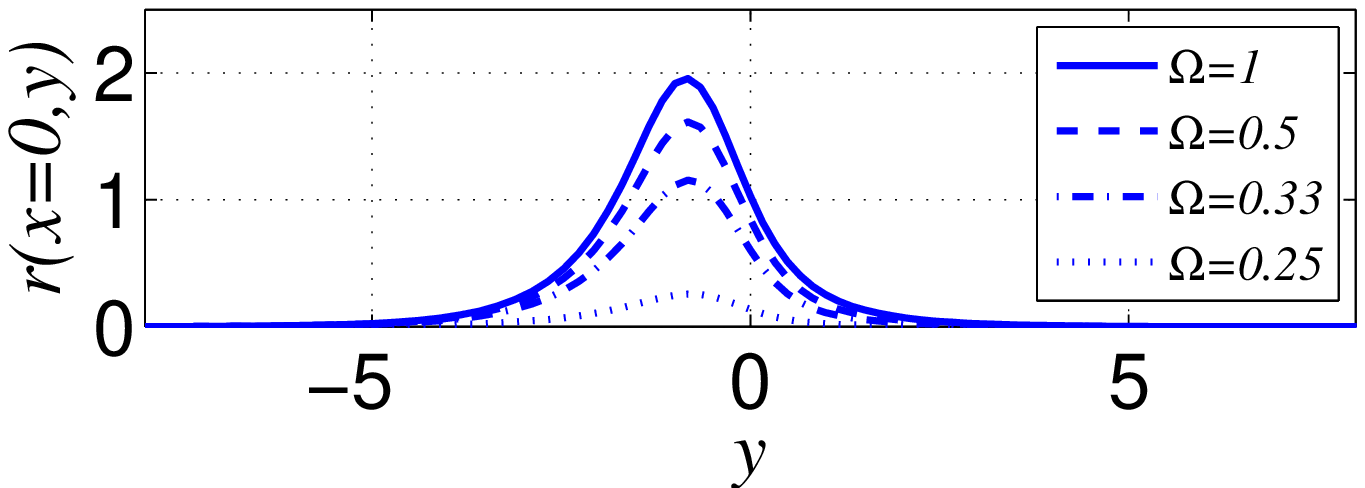}}\quad
  \put(-240,80){(a)}    \quad
  {\includegraphics[scale=0.6]{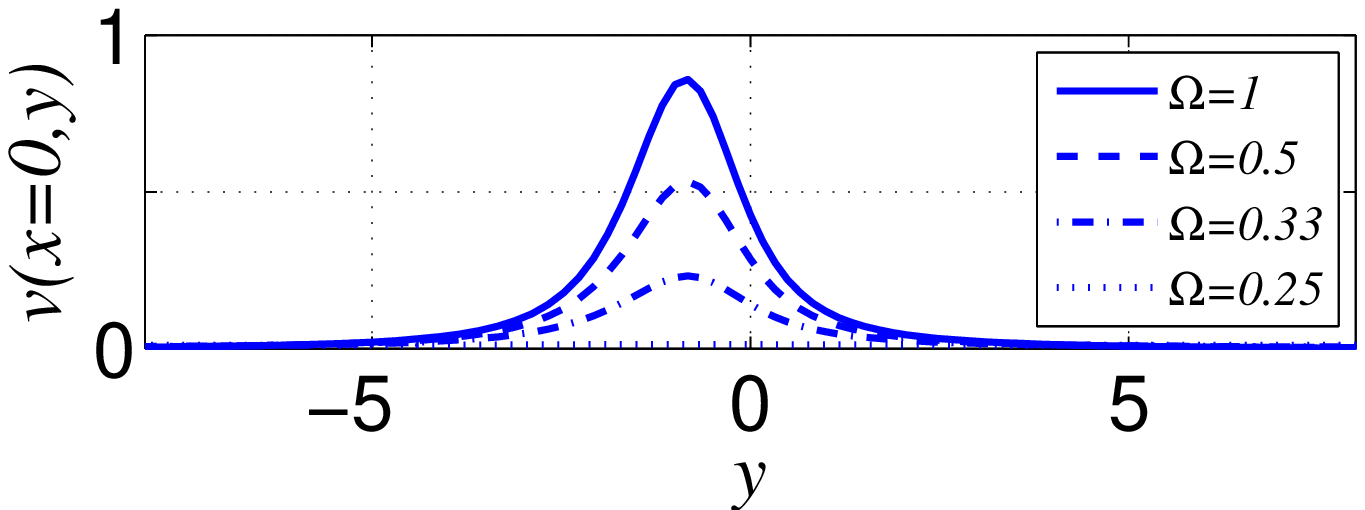}}\quad
  \put(-240,80){(b)}    \quad
	\caption{Cross sections of asymmetric flexural-gravity lumps for different oscillation frequencies $\Omega$ (introduced in Eq. \eqref{334}). (a),(b) respectively show cross sections $r(x=0,y)$, $v(x=0,y)$. Parameters used here are $\delta k=2.8,~\hat{R}=0.01, \hat{H}^{\f{1}{4}}=0.19$, $L=5$.}\label{case2}
\end{figure}

One important feature of the governing equation Eq. \eqref{940} in the presence of a non-zero thickness variation function is that the frequency of the periodic phase of the amplitude $\Omega$ appears explicitly in the equation (c.f. Eq. \eqref{942}, last term). In other words if $f(y)$=0, by the rescaling introduced $\Omega$ disappears from the governing equation Eq. \eqref{940}, but if $f(y)\neq$0 there is no rescaling that can remove $\Omega$. Therefore $\Omega$ appears as a free parameter in the governing equation. In the previous two examples (Figures \ref{case1}(a)-(f)), we set $\Omega$=1/3. Figure \ref{case2} shows the effect of $\Omega$ on the shape of the lump obtained ($f(y)$ and other parameters are the same as those in Figures \ref{case1}(a)-(c)). Clearly with the increase in $\Omega$ the height of the lump increases but the wavelength is almost unchanged. In the case presented in Figure \ref{case2}, for $\Omega<$0.24 the lump height vanished and we obtained a trivial solution of a constant mean current. Note for $\Omega=0.25$, the mean current also has hump-like profile but its amplitude is of O($0.01$).

\begin{figure}
\centering
  {\includegraphics[scale=0.6]{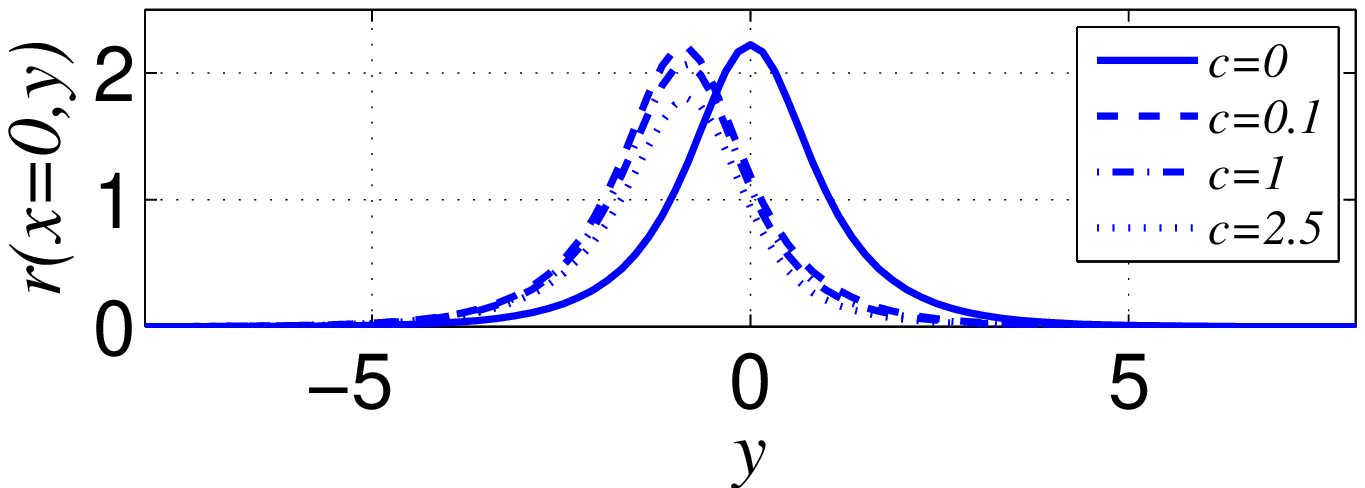}}\quad
  \put(-240,80){(a)}    \quad
  {\includegraphics[scale=0.6]{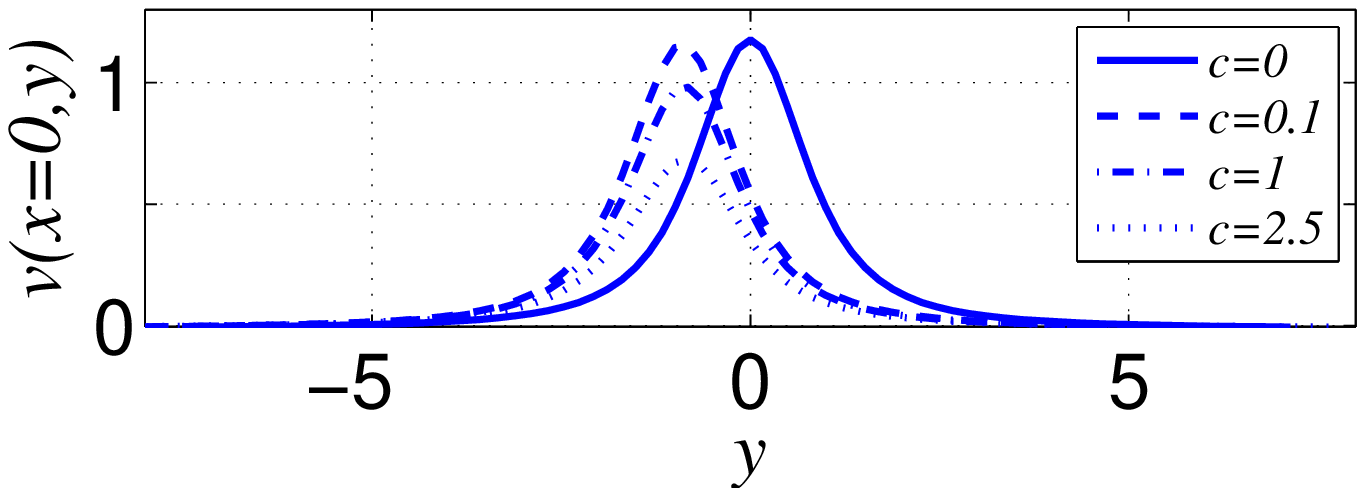}}\quad
  \put(-240,80){(b)}    \quad
	\caption{ Cross sections of asymmetric flexural-gravity lumps for different amplitudes of $f(y)$. (a),(b) show cross sections $r(x=0,y)$, $v(x=0,y)$ respectively. Parameters used here are $\delta k=2.8,~\hat{R}=0.01, \hat{H}^{\f{1}{4}}=0.19$, $L=5$.}\label{height}
\end{figure}

The effect of the amplitude of the thickness variation is higher on the lump height than the mean current. Figure \ref{height} compares lumps obtained for $f(y)=c\exp (-y^2) \sin y$ and for four choices of $c$=0, 0.1 ,1 and 2.5. The case $c$=0 corresponds to a uniform thickness function  (i.e. $f(y)$=0) that is known to yield symmetric lumps. Once variation function is introduced (i.e. $c>$0) the peak equilibrium location changes, but the peak stays at the same $y$ for different values of $c$. Further increase in the $c$ decreases both the height and mean current. Effect of the $c$ is small on the lump height ($\sim$ 15\% when $c$ changes by a factor of 25) and the mean current ($\sim$ 40\%). Clearly the mean current is affected more than the lump height by the height of the thickness anomaly .

For flexural-gravity waves, variable medium can be a result of variations in the water depth as well. Governing equation for flexural-gravity waves over a variable bottom can be derived via a similar procedure followed \S2 and the final equation is in the same form as in Eq. \eqref{930} with different coefficient $\nu_3$ (see Appendix). Similar procedure can be followed to derive the governing equation for Capillary-Gravity waves in the presence of bottom variations. Capillary-gravity lumps excited by a localized moving pressure have been observed in the laboratory\cite{Diorio2009,Diorio2011,Cho2011}. Our results suggest that carefully architected variable mediums may be used in laboratory experiments to trap lumps, or restrict their motion along desired paths.

\section{Conclusion}
Here we showed that asymmetric flexural-gravity lumps can exist on a surface of an inviscid and irrotational fluid covered by a (transversely) variable-thickness elastic material. Assuming that the variation in the thickness of the overlying elastic sheet is small, we derived, via perturbation expansion, the governing equation for the envelope of wavepackets in a flexural-gravity wave system. The governing equation is in the form of classical Davey-Stewartson equation but with an additional term in the surface evolution equation that accounts for the variation in the thickness. In order to find fully-localized solitary waves, we then carried out a continuation procedure by Lagrange interpolation, combined with Newton-Raphson iteration scheme. We showed that the peak of the asymmetric lump forms near a local minimum of the elastic sheet thickness. Also, in contrast to lumps propagating over a flat sheet, asymmetric lumps over a non-uniform thicknesses can only exist for frequencies greater than a minimum frequency. 

In practical applications the medium is often not perfectly uniform. In fact in many cases the variation is only in one spatial direction (e.g. sloping shoreline, sloping edge of the ice). Possibility of existence of lumps in these systems suggests that these fully-localized solitary waves may exist more widely than expected before. It also suggests where lumps may be more often expected: for instance, contours of constant water depth over a sloping beach (that has a local minimum along the beach slope) in ice-covered waters are loci of lumps. Whether these loci can act as attractors for wandering lumps is an interesting subject worth further research. 

\renewcommand{\appendixname}{APPENDIX}
\appendix*
\section{VARIABLE TOPOGRAPHY}

Here we present the governing equation for evolution of wavepackets of flexural-gravity waves, assuming that the thickness of the elastic sheet is constant, but the seabed has small variations in the transverse (to propagation) direction. 

We consider that the seabed is given by a mean bottom at the depth $h$, with small perturbations $b(y)/h\sim O(\epsilon^2)$. Under this assumption, governing equations are the same as Eq. \eqref{910} except the bottom boundary condition in Eq. \eqref{989} now is
\be
\phi_z=\phi_yb_y, ~~~z=b(y)\nn
\ee
We define a normalized bottom perturbation $b^*=bh/a^2$. Now using scaling variables we introduced before in Eq. \eqref{740}, the dimensionless bottom boundary condition Eq. \eqref{988}, after dropping asterisks, turns into
\be
\phi_z=\epsilon^2\delta^2\phi_yb_y, ~~~z=\epsilon^2 b(y)\nn
\ee
By substituting this into the governing equations and following the similar procedure we arrive at Eq. \eqref{930}, we in fact arrive at the same form equation with just a different coefficient $\nu_3$ which is now
\be
\nu_{3}=-b(y)\frac{\delta k \omega[-\delta c_{p}^{2} k \sinh \delta k+\cosh\delta k(\tilde{H}+1-\tilde{R}c_{p}^{2})]}{\sinh\delta k(\tilde{H}+1-\tilde{R}c_{p}^{2})+\delta c_{p}^{2}k \cosh\delta k+2\tilde{R}c_{p}^{2}\sinh\delta k}\nn
\ee

It can be shown that the sign of $\nu_3/b(y)$ is decided by the sign of $-(1+\tilde H)$, which is always negative. In the case of variable thickness the sign of $\nu_3/f(y)$ is decided by $\tilde{R}\tanh\delta k(2\tilde{H}-1)+3\tilde{H}\delta k$. If effect of inertia is small (i.e. $\tilde R\approx 0$), the sign of $\nu_3/f(y)$ is always positive. Since in the case of variable thickness lumps are formed near the local minima of the thickness variation function $f(y)$, we then conclude that in the case of variable seabed, lumps are formed near the local maxima of the seabed, i.e. where the water depth is minimum. Extension of these formula to the capillary-gravity waves over non-uniform water depth is also obtained, but is similar to the above case, and hence is not presented here.

\bibliography{ASYMMETRICLUMPS,518_Three_D_Soliton}
\end{document}

%% file: defs.tex
\usepackage{subeqnarray}
\usepackage{color}

\def\be{\begin{eqnarray}}
\def\ee{\end{eqnarray}}
\def\f{\frac}

\def\lb{\left[}
\def\rb{\right]}
\def\lcb{\left\{}
\def\rcb{\right\}}

\def\ep{\epsilon}

\def\befi{\begin{figure}}
\def\eefi{\end{figure}}

\def\bce{\begin{center}}
\def\ece{\end{center}} 
\def\nn{\nonumber}

\catcode`\!=\active
\def!#1!{}

\def\bes{\begin{subeqnarray}}
\def\ees{\end{subeqnarray}}